# Photochemical Synthesis of P-S-H Ternary Hydride at High Pressures

*Tingting Ye[1,2], Hong Zeng[1,2], Peng Cheng[1,2], Deyuan Yao[1,2], Xiaomei Pan[1,2], Xiao Zhang[1,3,]\*, Junfeng Ding[1,2,]\**

[1]Key Laboratory of Materials Physics, Institute of Solid State Physics, HFIPS, Chinese Academy of Sciences, Hefei 230031, China

[2]University of Science and Technology of China, Hefei 230026, China

[3]Frontiers Science Center for Transformative Molecules, School of Chemistry and Chemical Engineering, Shanghai Jiao Tong University, Shanghai 200240, China.

ABSTRACT: The recent discovery of room temperature superconductivity (283 K) in carbonaceous sulfur hydride (C-S-H) has attracted lots of interests in ternary hydrogen rich materials. In this report, ternary hydride P-S-H has been synthesized through photochemical reaction from elemental sulfur (S), phosphorus (P) and molecular hydrogen ($H_2$) at high pressures and room temperature. The Raman spectroscopy under pressure shows that $H_2S$ and $PH_3$ compounds are synthesized after laser heating at 0.9 GPa and a ternary van der Waals compound P-S-H is synthesized with a further compression to 4.6 GPa. The P-S-H compound is probably a mixed alloy of $PH_3$ and $(H_2S)_2H_2$ with a guest-host structure similar to the C-S-H system. The ternary hydride can persist up to 35.6 GPa at least and shows two phase transitions at approximately 23.6 GPa and 32.8 GPa, respectively. The P-S-H ternary hydride in this report is a competitive candidate for new hydride superconductors with near room-temperature transitions.

INTRODUCTION

Hydrides are materials in which hydrogen is combined with other elements to form ionic, covalent, or interstitial systems. They initially came to prominence because of their ability to reversibly store large amounts of hydrogen under moderate conditions.[1-4] Hydrides are also important in storage battery technologies such as nickel-metal hydride battery.[5-6] Recently, hydrides are promising materials for the realization of high temperature superconductivity as they can combine the unique prerequisites for superconductivity such as high-frequency phonons, strong electron-phonon coupling, and a high density of the electronic states.[7] Superconductivity in hydrides has been reinvigorated since discovery of a remarkably high superconducting transition temperature $T_C$ of 203 K at 150 GPa in $H_3S$ in 2015.[8] A lot of new hydrides are synthesized,[9-11] and record of the highest $T_C$ is broken by hydrides in rapid succession.[12-15]

To explore new hydrides with $T_C$ above room temperature, the investigation into the mechanism of the high $T_C$ of $H_3S$ have been a hot topic and fruitful.[16-19] It is suggested that superconductivity of $H_3S$ can be attributed to conventional phonon-mediated mechanism as it exhibits strong covalent bonds giving rise to large electron-phonon coupling.[16, 20] Increasing the covalent character of the sulfur-hydrogen bond by replacing S atoms with chalcogens or other atoms is a good choice to further enhance the superconducting properties of $H_3S$.[21] Several ternary hydrides based on the $H_3S$ structure are predicted to exhibit high-$T_C$ behavior.[21-24] Very recently, a carbonaceous sulfur hydride (C-S-H) synthesized through photochemical reaction at high pressure exhibits room-temperature superconductivity up to 283 K.[14] In addition, first-principles calculations predict a relative higher $T_C$ up to the 280 K via 2.5% P doping in $SH_3$ owing to P as a covalent atom lighter than S.[25] However, P-S-H system has never been observed in experiments, and its physical properties are still awaiting investigation.

The room-temperature superconductor C-S-H is synthesized by a photochemical method and has a molecular guest-host structure based on its Raman spectra before metallization.[14] A guest-host compound $(H_2S)_2H_2$ with the same stoichiometry as $H_3S$ is found at low pressures,[26] and transfer to superconducting $H_3S$ above 100 GPa.[27-28] The familiar way to synthesize $(H_2S)_2H_2$ is compressing $H_2S$ and $H_2$ together. This strategy is believed to be useful for synthesizing ternary guest-host compounds.[29]

Here, we report the synthesis of ternary van der Walls compound of P-S-H through photochemical reaction under pressure. For reference, $H_2S$, $PH_3$, and $PH_3$-$H_2$ samples are also synthesized and investigated by Raman spectroscopy. The pressure dependent Raman spectra suggest that the P-S-H compound is in a guest-host structure and has $H_2$ molecule with prolonged H-H distance. P-H and S-H modes are coexisting in the new hydride and are different from known compounds.

EXPERIMENTAL SECTION

Symmetric diamond anvil cells (DACs) equipped with anvils with central culets of 250 μm in diameter were employed in the experiments. Re gaskets were indented to 40 μm thickness and laser-drilled a hole in 110 μm diameter as sample chamber. Small pieces of S and red phosphorus (>99.99%) ordered from Sigma-Aldrich were positioned separately in the chamber and then filled with $H_2$ gas at ~200 MPa.

Because $H_2S$ will solidify at a low pressure about 1.1 GPa, in order to get fluid $H_2S$,[26] laser heating was performed at pressures of 0.9 GPa. 532 nm solid state laser at a power around 460 mW was used in laser heating to melt S and $P_{red}$. Laser sharply focused to a spot of 2 to 3 μm in diameter irradiate $P_{red}$ and S successively and move back and forth. Synthesis of $H_2S$ and $PH_3$ were judged from Raman spectra measured. After finishing synthesis of $H_2S$ and $PH_3$, the sample is compressed up to 36 GPa at room temperature in DAC.

Pressure was determined using the ruby fluorescence.[30] For the Raman experiments, a backscattering geometry was adopted for confocal measurements with incident laser wavelengths of 532 nm.[31-32] The Raman notch filters were of a very narrow bandpass (Optigrate) allowing Raman measurements down to 10 cm$^{-1}$ in the Stokes and anti-Stokes. One of these notch filters is used as a beam splitter to inject the laser into the optical path.

RESULTS AND DISCUSSION

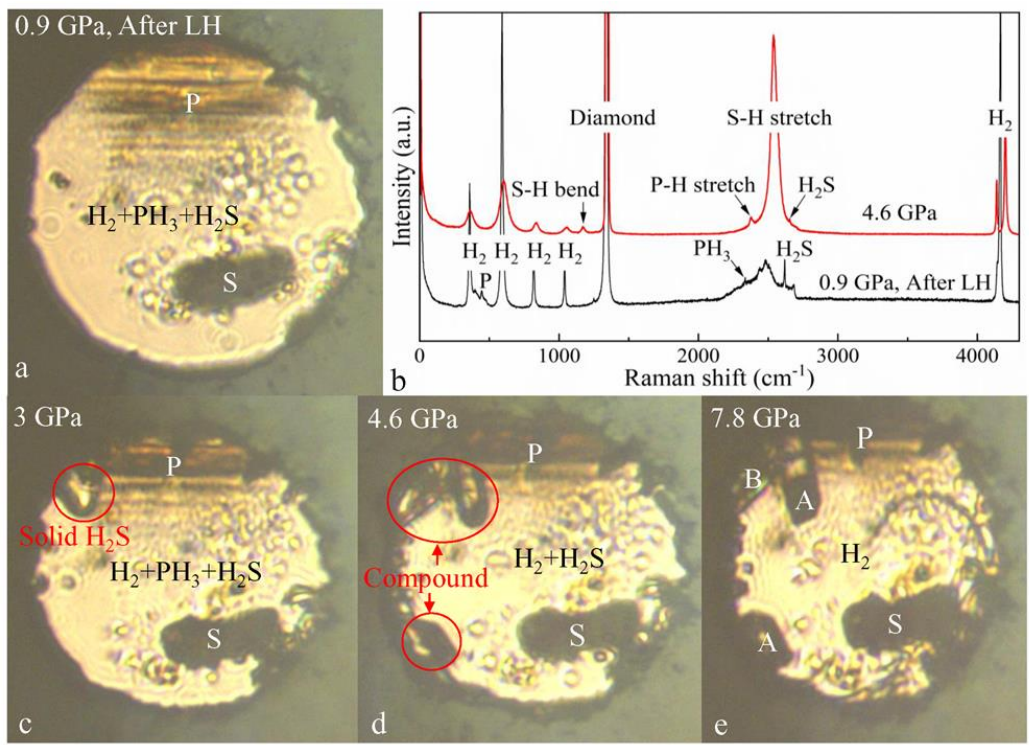

**Figure 1.** Micrographic images of sample at selected pressures (a) 0.9 GPa, (c) 3GPa, (d) 4.6 GPa, and (e) 7.8GPa. (b) Raman spectra at 0.9 GPa and 4.6 GPa.

After laser heating on red phosphorus ($P_{red}$) and $H_2$, $P_{red}$ is melted by laser obviously in visual observation. However, solid $P_{red}$ will generate back when we shot moderate laser on $H_2$ region to measure Raman spectrum. It can index that the synthesized compound of P and H is unstable and in fluid state. A new sharp Raman peak arises at about 2330 cm$^{-1}$ after we heat $P_{red}$ again for a long time (few minutes) using 460 mW laser, as shown in Figure 1b. This peak can be assigned to P-H stretching mode,[33] indicating synthesis of stable compound $PH_3$. In Figure 1a, $PH_3$ is fluid and only can be detected on and near P. On the other hand, $H_2S$ is easy to be obtained by heating S and $H_2$. $H_2S$ with the Raman peak at about 2620 cm$^{-1}$ fills the whole chamber.

In Figure 1c, $H_2S$ solidifies into a transparent crystal at 3 GPa. $PH_3$ still stays in fluid state at such pressures. At 4.6 GPa, solid $H_2S$ turns into black opaque and disappears quickly. Instead, some new crystals grow out near or attached on one side of gasket, as labeled in Figure 1d. They are all transparent but look like having lower transparency than solid $H_2S$. Raman spectrum of these new crystals in Figure 1b shows the coexistence of slightly broad peak of P-H stretching mode, high-intensity S-H stretching mode, and H-H vibron. Especially, the H-H vibron splits into two peaks at approximately 4200 cm$^{-1}$. One peak with the higher intensity is originated from pure $H_2$, and the other one has lower frequency implies the synthesis of $H_2$ molecule guest-host compound, such as $(H_2S)_2H_2$.[26] It is plausible that these new crystals are ternary compound composed of P, S, and H. P-H stretching mode is no longer detected at other positions, which suggests $PH_3$ all solidified into new crystals. On the other hand, some fluid $H_2S$ still exists in chamber as a high-frequency S-H mode can be detected everywhere. At 7.8 GPa, crystals change into two types according to virtual observation in Figure 1e. The first kind of crystal turns to black opaque (marked as A), and the other one is rather bright (marked as B).

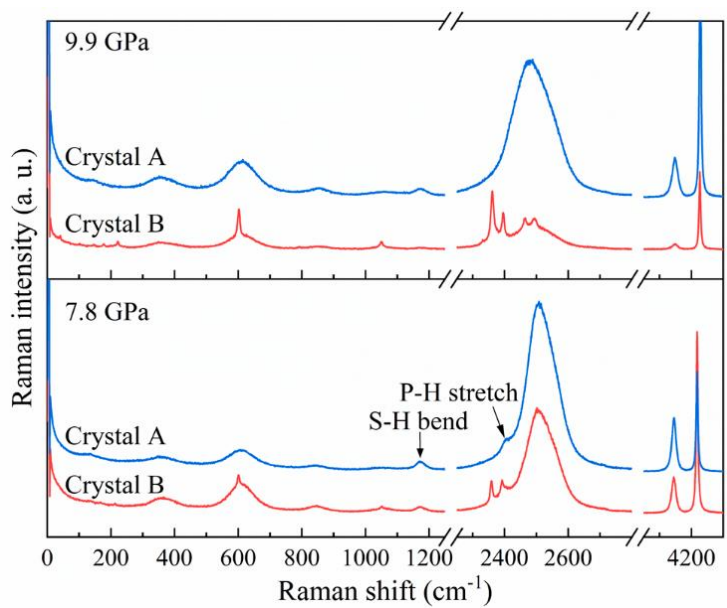

**Figure 2.** Raman spectra of the two crystals at 7.8 GPa and 9.9 GPa, respectively.

To further reveal the ingredient of the two kinds of crystal, Raman spectra of both crystals are investigated at different pressures as shown in Figure 2. At 7.8 GPa, the S-H stretching modes near 2500 cm$^{-1}$ and H-H vibron near 4200 cm$^{-1}$ of the two types of crystal are the same. Crystal A has only one weak peak of P-H stretch at about 2400 cm$^{-1}$ in Raman spectra, like a shoulder on S-H stretching mode. Crystal B has two clear peaks in P-H stretching region in Figure 2. Meanwhile, two sharp peaks appear at about 600 cm$^{-1}$ and 1050 cm$^{-1}$ in Crystal B, showing as prominences in roton mode of $H_2$. At 9.9 GPa in Figure 2, P-H stretching mode disappears in crystal A, and all the Raman peaks could be attributed to S-H stretching modes and H-H vibron, which indicates that crystal A is the van der Walls compound $(H_2S)_2H_2$ with molecular guest-host structure as reported in the earlier research of S-H compounds at high pressures.[26] For crystal B, asymmetric broad band of S-H stretching mode at around 2500 cm$^{-1}$ rises two clear peaks on top. The intensity of P-H stretching mode near 2400$^{-1}$ increases remarkably. Some weak peaks below 220 cm$^{-1}$ are originated from the lattice mode of $H_2S$, indicating orientational ordering occurs. The Raman spectra figure out transferring and gathering of P-H compound in crystal B coexisting with S-H and H-H bonds, which implies the formation of P-S-H ternary hydride.

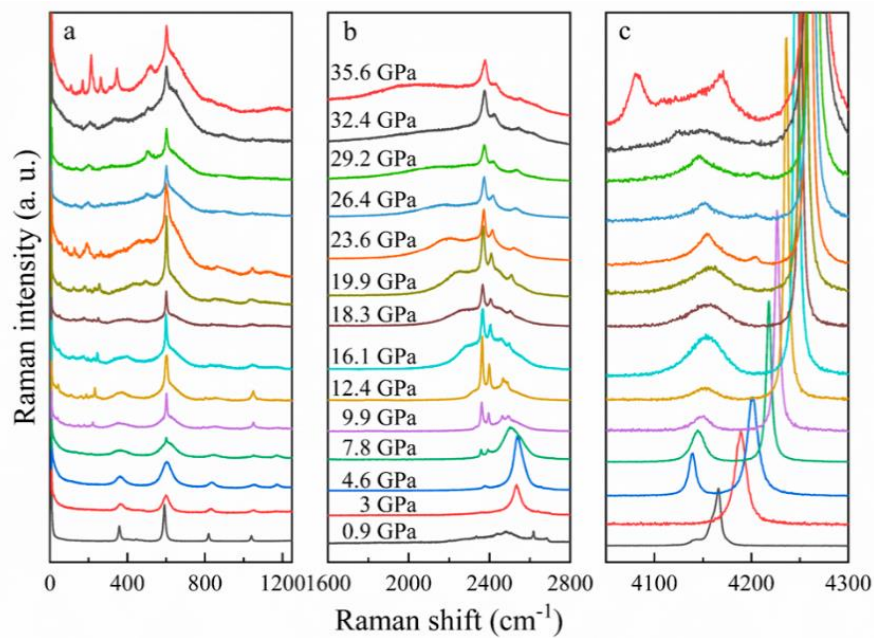

**Figure 3.** Raman spectra of crystal B at different pressures. (a) Lattice and bending modes, (b) stretching modes, and (c) H-H vibron.

In the following, let's focus on crystal B which is plausible to be a P-S-H ternary compound. The pressure dependencies of Raman spectra of crystal B up to 35.6 GPa are measured as shown in Figure 3. According to Raman active frequency of different bonds in literature,[14, 28, 34-36] the Raman spectra are divided into three regions, namely, the lattice and bending modes below 1200 cm$^{-1}$, the stretching modes of P-H and S-H between 1600 cm$^{-1}$ and 2800 cm$^{-1}$, and the H-H vibron above 4050 cm$^{-1}$. The origin of the Raman modes in crystal B are discussed in detail as follows.

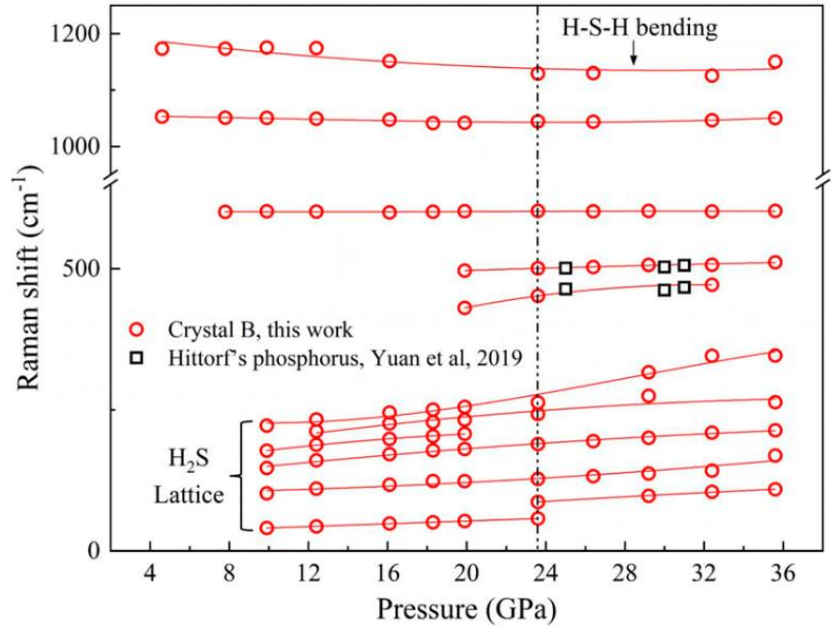

**Figure 4.** Pressure dependent frequencies of lattice and bending modes for crystal B. The solid lines are the guides to eyes and the. The roton modes of $H_2$ are not shown to be brevity. The vertical line shows the phase boundary.

In Figure 4 the lattice modes under 400 cm$^{-1}$ of crystal B could be attributed to $H_2S$ molecule.[14, 34] The highest-frequencies peak near 1200 cm$^{-1}$ can be distributed to H-S-H bending mode. Two peaks at 601 cm$^{-1}$ and 1050 cm$^{-1}$ are beyond example[37-38] and have extraordinarily little change with pressure. To our best knowledge, authentic attribution of these peaks is not found from the earlier literature, suggesting the synthesis of a new compound at 4.6 GPa. Upon compression, two new peaks near 400 cm$^{-1}$ and 500 cm$^{-1}$ appear at around 20 GPa because the sample decompose and generate Hittorf's phosphorus.[35] Meanwhile, $H_2S$ lattice modes show disappearance for peak at 180 cm$^{-1}$ and discontinuousness for peak at 30 cm$^{-1}$, which indicates a phase transition in crystal B near 23.6 GPa. The two new Raman modes in Figure 4 suggests the synthesis of a new compound differing from H-S or P-H compounds at 4.6 GPa.

Pressure dependencies of stretch modes for crystal B are shown in Figure 5. For reference, $H_2S$, $PH_3$, and $PH_3$-$H_2$ samples are also synthesized individually by photochemical method. The Raman modes of crystal B under pressure are compared with that of $H_2S$, $PH_3$, and $PH_3$-$H_2$ samples and analyzed as follows. First of all, the dotted lines in Figure 5 show that four peaks of S-H stretching modes in crystal B is similar to $(H_2S)_2H_2$,[28] which suggests that Crystal B adopts the framework of $(H_2S)_2H_2$. For the second, the two Raman peaks at 2300 cm$^{-1}$ and 2350 cm$^{-1}$ are in correspondence with $PH_3$[35] or $(PH_3)_2H_2$[36] at low pressures, figuring out that these two peaks could be attributed to P-H stretching mode. Another Raman peak at 2380 cm$^{-1}$ is in good accord with the earlier report of $(PH_3)_2H_2$ at 4.6 GPa.[36] As $(PH_3)_2H_2$ is insufficiently studied by Raman spectroscopy for pressures above 7 GPa in literature, we synthesis $PH_3$-$H_2$ samples and investigate its Raman spectra in detail to further reveal the origin of the Raman peaks for crystal B at higher pressures, as shown in Figure 5. At low pressure region below 7 GPa, P-H bond of the reference sample well match the earlier reports of $(PH_3)_2H_2$ (squares in Figure 5), indicating good sample quality and authentic experimental setup. The pressure dependencies of P-H stretching mode of crystal B are similar to that of the reference $(PH_3)_2H_2$ sample at low pressure region and show notable deviation above 7 GPa. Thus, the three peaks between 2300 cm$^{-1}$ to 2450 cm$^{-1}$ of crystal B belong to P-H stretching mode but are not originated from $(PH_3)_2H_2$ or $PH_3$. Thirdly, the other three peaks (circles with solid lines) diverge from any known S-H or P-H compounds, which indicates that crystal B is not a mixture of S-H and P-H compounds. Two phase transitions could be defined according to the disappearance and appearance of Raman peaks, one at around 23.6 GPa and the other one at around 32.8 GPa. The stretching modes in Figure 5 suggest Crystal B is a van der Waals compound that contains $PH_3$ molecular.

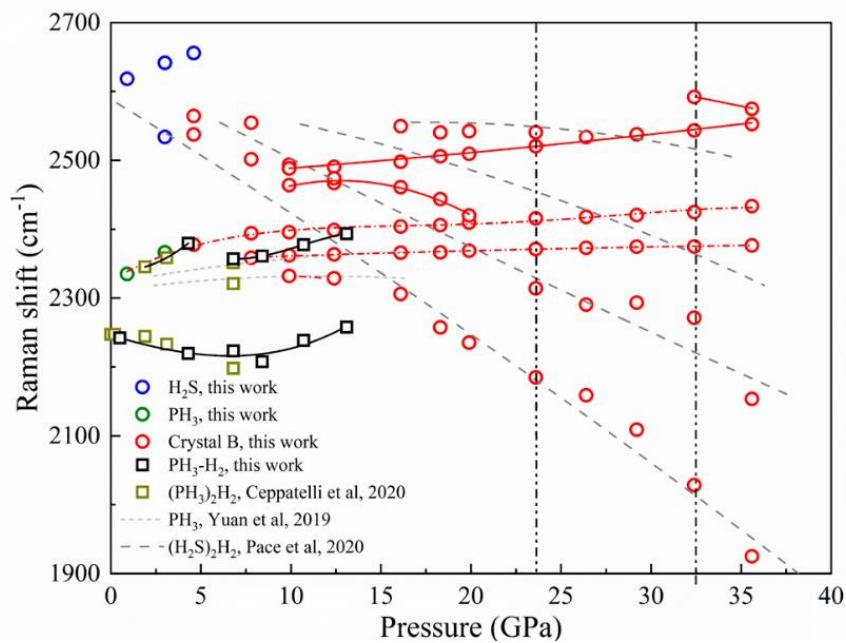

**Figure 5.** Pressure dependencies of P-H and S-H stretching modes for crystal B, $PH_3$, $H_2S$ and $PH_3$-$H_2$. Raman frequencies of $PH_3$[35], $(PH_3)_2H_2$[36], and $(H_2S)_2H_2$[28] are also shown as reference. The solid and dash dot lines are the guides to eyes. The vertical lines show the phase boundaries.

The H-H vibron modes in crystal B are consistent with $(H_2S)_2H_2$[28] in Figure 6, indicating that crystal B has $(H_2S)_2H_2$ as framework. We notice that there are still some small distinctions between crystal B and $(H_2S)_2H_2$. The first splitting of H-H vibron arises at 23.6 GPa for crystal B, higher than 16.7 GPa for $(H_2S)_2H_2$. In addition, two vibron modes at approximately 4140 cm$^{-1}$ and 4170 cm$^{-1}$ for $(H_2S)_2H_2$ can not be observed in crystal B. The discrepancies are probably originated from the insertion of $PH_3$ molecular in $(H_2S)_2H_2$ for crystal B. On the other hand, the H-H vibron of $(PH_3)_2H_2$ does not exist in crystal B. Two phase transition pressures determined from the H-H vibron are consistent with the values 23.6 GPa and 32.8 GPa in Figure 5. The H-H vibron in crystal B supports that the ternary compound is a mixed alloy with the $PH_3$ molecule inserted into $(H_2S)_2H_2$ framework.

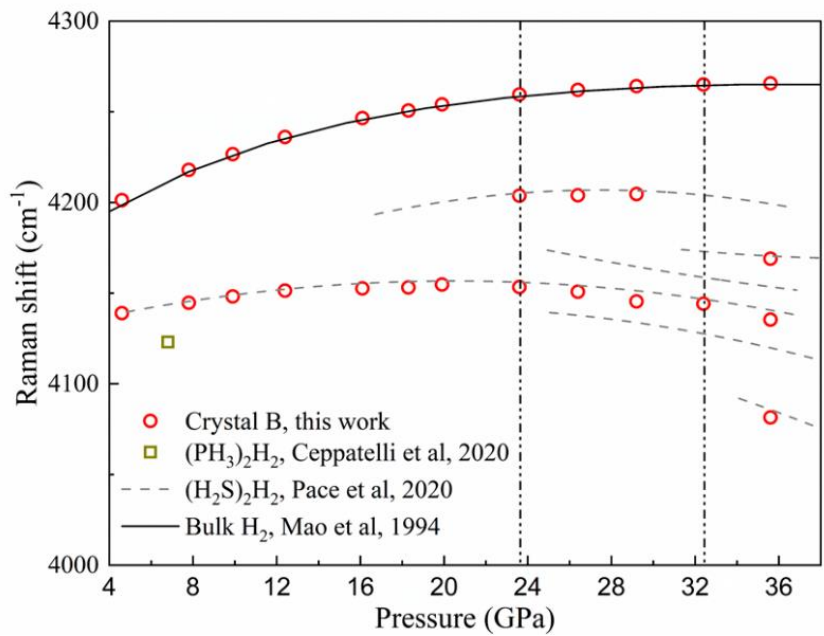

**Figure 6.** Pressure dependencies of H-H vibron for crystal B. The vertical lines show the phase boundaries.

According to the pressure dependencies of Raman modes, three features of crystal B are listed as follows. Firstly, crystal B is a new compound formed at 4.6 GPa, which can persist up to 35.6 GPa at least with two phase transitions under pressure. Secondly, P-H stretching modes in crystal B are similar to that in $PH_3$ or $(PH_3)_2H_2$ at low pressure region and differ above 7 GPa, suggesting crystal B contains $PH_3$ molecular. Thirdly, crystal B adopts the framework of $(H_2S)_2H_2$ with notable differences in H-H vibron above 23.6 GPa. Thus, crystal B is very likely to be a P-S-H ternary hydride with the guest-host structure, namely, $PH_3$ molecule inserts into the $(H_2S)_2H_2$ host lattice as a guest similar to the room-temperature superconductor C-S-H.[14] Accurate structural analysis is always an extremely challenging for hydride superconductors, because the light elements in hydrides lead to a very weak signal in X-ray scattering techniques.[8, 12, 14, 39] Raman spectroscopy is a powerful and frequently-used tool to probe the chemical and structural transformations of hydrides under pressure.[40-41] In fact, the crystal structure for the well-known C-S-H is still in debt. Theoretical calculations have hypothesized several different crystal structures for C-S-H.[14, 42] To further reveal the structural properties of the P-S-H ternary hydride, synchrotron x-ray diffraction and theoretical analysis are needed in the future.

The guest-host structures for both $(H_2S)_2H_2$ and C-S-H hydrides at low pressures become the building blocks of superconducting compounds with strong covalent character at high pressures, which exhibit high transition temperature superconductivity driven by strong electron-phonon coupling to high-frequency hydrogen phonon modes.[14,43-44] The earlier theoretical study has prophesied that the P-S-H ternary hydride shows superconducting transition temperature $T_C$ up to the 280 K with a cubic structure in $Im\bar{3}m$ phase.[25] Although the P-S-H in this report is a van der Waals compound with guest-host structure, the new ternary hydride is also plausible to transfer into a covalent compound through atomic substitution at high pressures. Thus, the synthesized P-S-H ternary hydride here with a guest-host structure similar to $(H_2S)_2H_2$ and C-S-H at low pressures is a competitive candidate for near room-temperature superconductor.

CONCLUSIONS

In conclusion, P-S-H ternary hydride has been synthesized at high pressures, and the Raman spectra are investigated in detail. The coexistence of P-H stretching mode, S-H stretching mode and H-H vibrion suggest the formation of P-S-H compound at 4.8 GPa. Pressure dependencies of the Raman mode up to 35.6 GPa indicate the ternary hydride is a van der Waals compound with a guest-host structure similar to the room-temperature superconductor C-S-H. Two phase transitions appear respectively at around 23.6 GPa and 32.8 GPa for the P-S-H ternary hydride.

The synthesized P-S-H ternary compound shed light on seeking after new hydride superconductors with near room-temperature transitions.

AUTHOR INFORMATION

**Corresponding Author**

*Xiao Zhang − *Key Laboratory of Materials Physics, Institute of Solid State Physics, HFIPS, Chinese Academy of Sciences, Hefei 230031, China; Frontiers Science Center for Transformative Molecules, School of Chemistry and Chemical Engineering, Shanghai Jiao Tong University, Shanghai 200240, China;*

Email: xiaozhang@issp.ac.cn.

*Junfeng Ding−*Key Laboratory of Materials Physics, Institute of Solid State Physics, HFIPS, Chinese Academy of Sciences, Hefei 230031, China; University of Science and Technology of China, Hefei 230026, China;*

Email: junfengding@issp.ac.cn.

**Author Contributions**

The manuscript was written through contributions of all authors. All authors have given approval to the final version of the manuscript.

**Notes**

The authors declare no competing financial interest.

ACKNOWLEDGMENT

This work was supported by National Natural Science Foundation of China (Grant Nos. 52002372, 51672279, 51727806, 11874361, and 11774354), Science Challenge Project (No. TZ2016001), and CAS Innovation Grant (No. CXJJ-19-B08).

Pressures. *Nature* **2019**, *569*, 528-531.
(13) Somayazulu, M.; Ahart, M.; Mishra, A. K.; Geballe, Z. M.; Baldini, M.; Meng, Y.; Struzhkin, V. V.; Hemley, R. J. Evidence for Superconductivity above 260 K in Lanthanum Superhydride at Megabar Pressures. *Physical Review Letters* **2019**, *122*, 027001.
(14) Snider, E.; Dasenbrock-Gammon, N.; McBride, R.; Debessai, M.; Vindana, H.; Vencatasamy, K.; Lawler, K. V.; Salamat, A.; Dias, R. P. Room-Temperature Superconductivity in a Carbonaceous Sulfur Hydride. *Nature* **2020**, *586*, 373-377.
(15) Cheng, Y.; Wang, X.; Zhang, J.; Yang, K.; Niu, C.; Zeng, Z. Superconductivity of Boron-Doped Graphane under High Pressure. *RSC Advances* **2019**, *9*, 7680-7686.
(16) Errea, I.; Calandra, M.; Pickard, C. J.; Nelson, J.; Needs, R. J.; Li, Y.; Liu, H.; Zhang, Y.; Ma, Y.; Mauri, F. High-Pressure Hydrogen Sulfide from First Principles: A Strongly Anharmonic Phonon-Mediated Superconductor. *Physical Review Letters* **2015**, *114*, 157004.
(17) Komelj, M.; Krakauer, H. Electron-Phonon Coupling and Exchange-Correlation Effects in Superconducting $H_3S$ under High Pressure. *Physical Review B* **2015**, *92*, 205125.1-205125.5.
(18) Einaga, M.; Sakata, M.; Ishikawa, T.; Shimizu, K.; Eremets, M. I.; Drozdov, A. P.; Troyan, I. A.; Hirao, N.; Ohishi, Y. Crystal Structure of the Superconducting Phase of Sulfur Hydride. *Nature physics* **2016**, *12*, 835-838.
(19) Bernstein, N.; Hellberg, C. S.; Johannes, M. D.; Mazin, I. I.; Mehl, M. J. What Superconducts in Sulfur Hydrides under Pressure and Why. *Physical Review B* **2015**, *91*, 060508.
(20) Errea, I.; Calandra, M.; Pickard, C. J.; Nelson, J. R.; Needs, R. J.; Li, Y.; Liu, H.; Zhang, Y.; Ma, Y.; Mauri, F. Quantum Hydrogen-Bond Symmetrization in the Superconducting Hydrogen Sulfide System. *Nature* **2016**, *532*, 81-84.
(21) Heil, C.; Boeri, L. Influence of Bonding on Superconductivity in High-Pressure Hydrides. *Physical Review B* **2015**, *92*, 060508.
(22) Liu, B. B.; Cui, W. W.; Shi, J. M.; Zhu, L.; Chen, J.; Lin, S. Y.; Su, R. M.; Ma, J. Y.; Yang, K.; Xu, M. L.; Hao, J.; Durajski, A. P.; Qi, J. S.; Li, Y. L.; Li, Y. W. Effect of Covalent Bonding on the Superconducting Critical Temperature of the H-S-Se System. *Physical Review B* **2018**, *98*, 8.
(23) Amsler, M. Thermodynamics and Superconductivity of $S_xSe_{1-x}H_3$. *Physical Review B* **2019**, *99*, 060102.
(24) Chang, P. H.; Silayi, S.; Papaconstantopoulos, D. A.; Mehl, M. J. Pressure-Induced High-Temperature Superconductivity in Hypothetical $H_3X$ (X=As, Se, Br, Sb, Te and I) in the $H_3S$ Structure with $Im\bar{3}M$ Symmetry. *Journal of Physics and Chemistry of Solids* **2020**, *139*, 109315.
(25) Ge, Y. F.; Zhang, F.; Yao, Y. G. First-Principles Demonstration of Superconductivity at 280 K in Hydrogen Sulfide with Low Phosphorus Substitution. *Physical Review B* **2016**, *93*, 224513.
(26) Strobel, T. A.; Ganesh, P.; Somayazulu, M.; Kent, P. R.; Hemley, R. J. Novel Cooperative Interactions and Structural Ordering in $H_2S-H_2$. *Phys Rev Lett* **2011**, *107*, 255503.
(27) Guigue, B.; Marizy, A.; Loubeyre, P. Direct Synthesis of Pure $H_3S$ from S and H Elements: No Evidence of the Cubic Superconducting Phase up to 160 Gpa. *Physical Review B* **2017**, *95*, 020104.
(28) Pace, E. J.; Liu, X. D.; Dalladay-Simpson, P.; Binns, J.; Peña-Alvarez, M.; Attfield, J. P.; Howie, R. T.; Gregoryanz, E. Properties and Phase Diagram of $(H_2S)_2H_2$. *Physical Review B* **2020**, *101*, 174511.
(29) Bykova, E.; Bykov, M.; Chariton, S.; Prakapenka, V. B.; Glazyrin, K.; Aslandukov, A.; Aslandukova, A.; Criniti, G.; Kurnosov, A.; Goncharov, A. F. Structure and Composition of C-S-H Compounds up to 143 Gpa. *Physical Review B* **2021**, *103*, L140105.
16